\documentclass[pra,twocolumn,10pt,aps,superscriptaddress,longbibliography]{revtex4-2}  
\setcitestyle{square}
\usepackage{graphicx}  
\usepackage{float}
\usepackage{dcolumn}   
\usepackage{bm}        
\usepackage{amssymb}   
\usepackage{amsmath}
\usepackage{isomath}
\usepackage[usenames,dvipsnames]{color}
\usepackage{soul}
\usepackage{mathtools}
\usepackage{physics}
\usepackage[normalem]{ulem} 
\usepackage{xpatch}
\usepackage{IEEEtrantools}
\usepackage[newcommands]{ragged2e}
\usepackage{subcaption}
\usepackage{pgffor}
\usepackage{xprintlen}
\usepackage{comment}
\usepackage{parskip}
\usepackage{lipsum} 
\usepackage{braket}
\usepackage{dirtytalk}

\newcommand{\GT}[1]{\noindent \color{green}(GT: #1)\normalcolor}

\usepackage{xcolor}
\usepackage{hyperref}
\hypersetup{
     colorlinks  = true,
     linkcolor    = red,
     citecolor    = cyan,
     urlcolor     = magenta
     }

\begin{document}
\title{Edge States with Hidden Topology in Spinner Lattices}
\author{Udbhav Vishwakarma}
\affiliation{Department of Aerospace Engineering, Indian Institute of Science, Bangalore 560012, India}

\author{Murthaza Irfan}
\affiliation{Department of Aerospace Engineering, Indian Institute of Science, Bangalore 560012, India}

\author{Georgios Theocharis}
\affiliation{LAUM, CNRS-UMR 6613, Le Mans Universit\'{e}, Avenue Olivier Messiaen, 72085 Le Mans, France}

\author{Rajesh Chaunsali}%
\email{Corresponding author: rchaunsali@iisc.ac.in}
\affiliation{Department of Aerospace Engineering, Indian Institute of Science, Bangalore 560012, India}%

\date{\today}

\begin{abstract}
Symmetries---whether explicit, latent, or hidden---are fundamental to understanding topological materials. This work introduces a prototypical spring-mass model that extends beyond established canonical models, revealing topological edge states with distinct profiles at opposite edges. These edge states originate from hidden symmetries that become apparent only in deformation coordinates, as opposed to the conventional displacement coordinates used for bulk-boundary correspondence. Our model, realized through the intricate connectivity of a spinner chain, demonstrates experimentally distinct edge states at opposite ends. By extending this framework to two dimensions, we explore the conditions required for such edge waves and their hidden symmetry in deformation coordinates. We also show that these edge states are robust against disorders that respect the hidden symmetry. This research paves the way for advanced material designs with tailored boundary conditions and edge state profiles, offering potential applications in fields such as photonics, acoustics, and mechanical metamaterials.
\end{abstract}

\maketitle  


\textcolor{red}{\textbf{INTRODUCTION}}

Symmetry is a fundamental principle across various branches of physics, playing a significant role in the physics of topological materials and enabling robust, defect-immune manipulation of electrons~\cite{kane2005z,fu2007topological,fu2011topological}. These underlying symmetries have not only advanced our understanding of electronic properties but have also inspired the design of topological architectures for bosons, leading to desirable mechanical~\cite{prodan2009topological,kane2014topological} and optical properties~\cite{haldane2008possible,bahat2008symmetry,raghu2008analogs,khanikaev2013photonic}.

Toy models in one dimension, such as bipartite tight binding and stiffness or mass dimers, have long served as versatile platforms for exploring topological physics grounded in such symmetries. Notable examples include the Su-Schrieffer-Heeger (SSH) model~\cite{SSH1980}, Maxwell lattices~\cite{kane2014topological,mao2018maxwell}, and dimer configurations in one dimension~\cite{shi2021disorder}, which belong to the BDI topological class and exhibit chiral, particle-hole, and time-reversal symmetries~\cite{kitaev2009periodic,susstrunk2016classification}. According to the principle of bulk-boundary correspondence, edge states can appear in finite structures if the bulk is topologically non-trivial and the boundaries preserve the underlying symmetry~\cite{hasan2010colloquium,asboth2016short}. These models have not only provided foundational insights but have also inspired the creation of complex topological materials in various dimensions~\cite{Haldane1988,kane2005z,fu2007topological,fu2011topological,benalcazar2017quantized,liu2017novel,xie2019visualization,zhang2019dimensional,zheng2020three,zhang2023flexible}.

Recent research has unveiled that symmetries can also manifest in latent~\cite{zheng2023robust,rontgen2024topological} or hidden forms~\cite{andrianov1997matrix,hou2013hidden,li2015hidden,danawe2022finite,allein2023strain}. For instance, in spring-mass models—widely applicable across mechanical, optical, acoustic systems, and electrical circuits—a mass dimer chain with an odd number of two altering masses and uniform stiffness, and with free boundaries is known to support edge states~\cite{wallis1957effect,deymier2013acoustic}. However, the topological origin of these states was not well understood. Allein et al.~\cite{allein2023strain} demonstrated that the underlying chiral symmetry for topological characterization emerges in deformation coordinates rather than the traditional displacement coordinates used for spring-mass models. This insight opens new avenues for designing architectured materials by considering varying masses which could support novel topological states protected by hidden chiral symmetry.

In this work, we enlarge the design space of topological lattices by varying mass along with different types of spring connections within the unit cell. 
We show that by using complex spring connections~\cite{susstrunk2015observation,pal2016helical,matlack2018designing,serra2018observation,zhu2020distinguishing,fang2023dispersion}, mass dimer 1D and 2D lattices could support distinct and unconventional profiles of topologically robust edge states at opposite ends---a phenomenon not observed in canonical two-band models~\cite{pal2017edge, Chen2018JMPS, Palmer2022NJP, allein2023strain, liu2023classical}.
Such profiles result from the breakage of both chiral and mirror symmetries. However, we show that the edge states are topological since both chiral and mirror symmetries are revealed in deformation coordinates.

Furthermore, we propose a chain of spinners strategically interconnected by springs as a physical realization of this model. We experimentally demonstrate different edge states at opposite ends, with their topological nature revealed through deformation coordinate analysis. We also extend this framework to two dimensions, highlighting the intricate fine-tuning required to observe the hidden symmetry of propagating waves on opposite edges of the lattice. Additionally, we perform a disorder analysis based on the underlying hidden symmetry, showing that symmetry-protected disorder can be leveraged to develop material architectures supporting robust topological edge states.

This work enhances our understanding of hidden symmetries in topological systems and opens new pathways for designing advanced materials with tailored edge state profiles, potentially impacting fields such as photonics, acoustics, and mechanical metamaterials.

\textcolor{red}{\textbf{RESULTS AND DISCUSSION}}

\textbf{Dimer models}

\begin{figure*}[t!]
    \centering
    \includegraphics[width=1.0\textwidth]{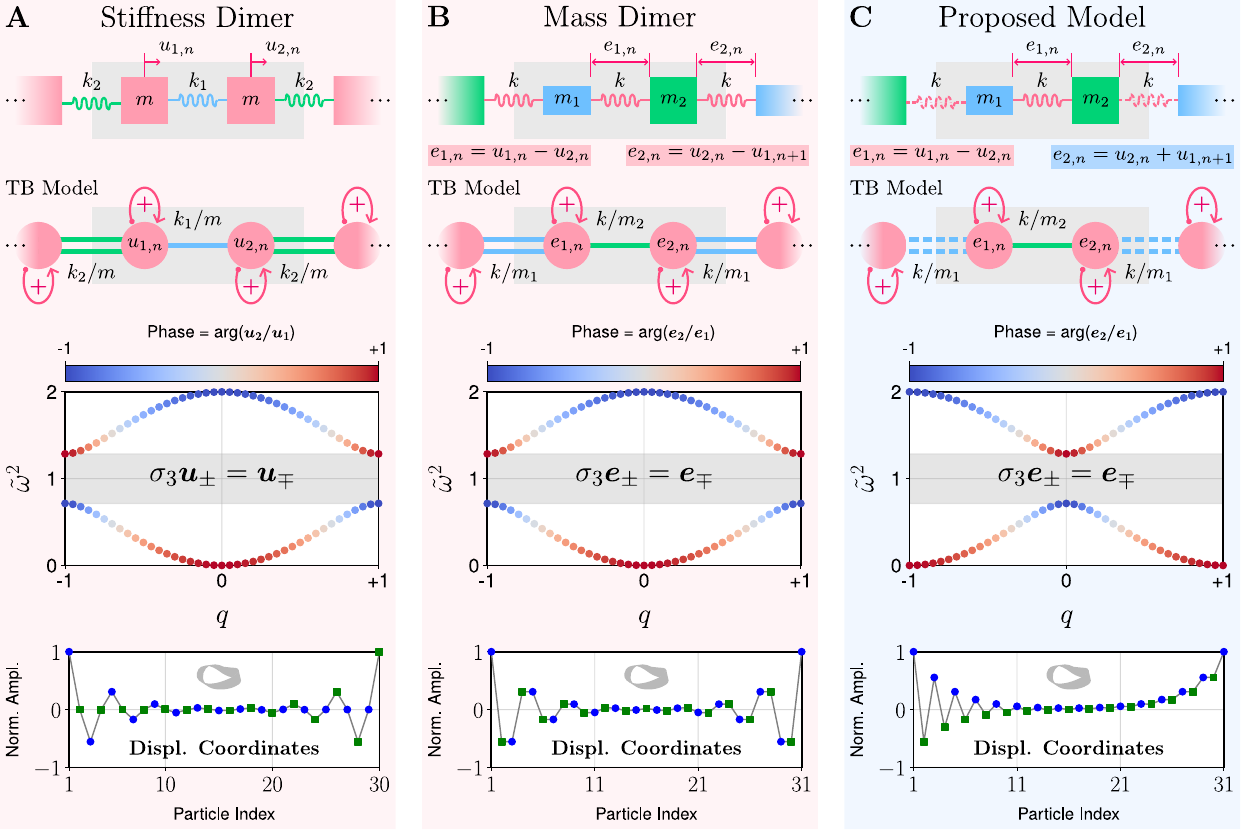}
    \caption{\textbf{Deformation mass dimer} --
    \textbf{(A)} Stiffness dimer with alternating spring constants and uniform masses and its tight-binding equivalent below. The bulk spectrum has chiral symmetry. Boundary states with the same localization profiles on both ends for topologically non-trivial finite lattices with mirror symmetry. 
    \textbf{(B)} Mass dimer with alternating masses but with uniform spring constants and the tight-binding equivalent that maps to the deformation coordinates. The bulk spectrum exhibits chiral symmetry only in deformation coordinates. Boundary states with the same localization profiles on both ends for topologically non-trivial finite lattices with mirror symmetry. 
    \textbf{(C)} The proposed model is a deformation mass dimer that resembles a mass dimer but with two different spring connections. Solid spring stores the potential energy in a usual way, but the dashed spring stores the potential energy differently due to the different nature of deformation. The tight-binding equivalent of the deformation-based model has positive (solid) and negative (dashed) shopping. The bulk spectrum exhibits chiral symmetry in deformation coordinates. Furthermore, a topologically non-trivial finite lattice without mirror symmetry supports mid-gap edge states with different localization profiles at the left and right ends.
    The shaded region in the lattices denotes the unit cell.}
    \label{Fig1}
\end{figure*}
In Fig.~\ref{Fig1}, we illustrate known spring-mass dimer models alongside the proposed model, which leads to various types of topological edge states. Figure~\ref{Fig1}A depicts the canonical 1D stiffness dimer comprising uniform masses but linked with two alternating spring constants $k_1$ and $k_2$. In the displacement framework, the equations of motion for the $n$th unit cell are given as
\begin{equation}
    \begin{aligned}
        m \Ddot{u}_{1,n} &= -k_1 (u_{1,n} - u_{2,n}) + k_2(u_{2,n-1} - u_{1,n}) \\
        m \Ddot{u}_{2,n} &= +k_1 (u_{1,n} - u_{2,n}) - k_2(u_{2,n} - u_{1,n+1}),
    \end{aligned}
    \label{SDDimer_EqsMotion}
\end{equation}
where $u_{1,n}$ and $u_{2,n}$ represent the displacements of the first and second particle in the $n$th unit cell. Upon substituting plane wave solutions, we get the non-dimensional bulk displacement dynamical matrix $\mathcal{D}_u(P,q)$ that can be represented in the Pauli basis as follows: ${\mathcal{D}_u(P,q) = c_0 \mathcal{I} + c_1 \sigma_1 + c_2 \sigma_2 + c_3 \sigma_3}$, where $c_0 = 1+P$, $c_1 = -(1+P \cos{q})$, $c_2 = -P \sin{q}$, ${c_3 = 0}$, $P \coloneqq k_2/k_1$, and $q \in [-\pi,\pi]$ is the wavenumber. 
$\mathcal{D}_u(P,q)$ possesses 
chiral symmetry, i.e., ${\sigma_3 \hat{\mathcal{D}}_u(P,q) \sigma_3 = -\hat{\mathcal{D}}_u(P,q)}$, where $\hat{\mathcal{D}}_u(P,q)$ is the trace-less counterpart. 
The dynamical matrix maps to the Hamiltonian of the standard SSH chain with two different hopping strengths and an onsite potential. Consequently, the spectrum is symmetric about a finite frequency, and the eigenvectors $u_{\pm}$ of the acoustic and optical bands form chiral pairs. The system has a well-defined topological invariant, namely, a winding number. A nonzero winding for $P>1$ indicates non-trivial topology, and a corresponding finite lattice with even particles and fixed edges preserves chiral symmetry and supports boundary states on the left and right ends. Since the finite lattice preserves the mirror symmetry, both boundary states also have the same localization profiles.

Figure~\ref{Fig1}B depicts the canonical 1D mass dimer comprising links with uniform spring constants but alternating masses, $m_1$ and $m_2$. In the displacement framework, the equations of motion for the $n$th unit cell are given as
\begin{equation}
    \begin{aligned}
        m_1 \Ddot{u}_{1,n} &= -k (u_{1,n} - u_{2,n}) + k(u_{2,n-1} - u_{1,n}) \\
        m_2 \Ddot{u}_{2,n} &= +k (u_{1,n} - u_{2,n}) - k(u_{2,n} - u_{1,n+1}).
    \end{aligned}
    \label{MDDimer_EqsMotion}
\end{equation}
Upon substituting plane wave solutions, we get the non-dimensional bulk displacement dynamical matrix $\mathcal{D}_u(P,q)$ that has the following Pauli components: ${c_0 = 1+1/P}$, ${c_1 = -(1+ \cos{q})/\sqrt{P}}$, $c_2 = - \sin{q}/\sqrt{P}$, and ${c_3 = 1-1/P}$, where $P \coloneqq m_2/m_1$. The bulk displacement dynamical matrix does not exhibit chiral symmetry. Therefore, topological characterization is ambiguous. However, a finite lattice with odd particles,  free boundaries, and lighter masses at the ends supports boundary states~\cite{wallis1957effect}. Interestingly, the system has topological features that are revealed by considering spring deformations as degrees of freedom. These were referred to as ``strain" coordinates in~\cite{allein2023strain}. In the deformation framework, the equations of motion for the $n$th unit cell are given as
\begin{equation}
    \begin{aligned}
        \frac{1}{k} \Ddot{e}_{1,n} &= -\frac{1}{m_2} (e_{1,n} - e_{2,n}) + \frac{1}{m_1} (e_{2,n-1} - e_{1,n}) \\
        \frac{1}{k} \Ddot{e}_{2,n} &= +\frac{1}{m_2} (e_{1,n} - e_{2,n}) - \frac{1}{m_1} (e_{2,n} - e_{1,n+1}),
    \end{aligned}
    \label{MDDimer_EqsMotion_Elong}
\end{equation}
where $e_{1,n} \coloneqq u_{1,n} - u_{2,n}$ and $e_{2,n} \coloneqq u_{2,n} - u_{1,n+1}$ represent the spring deformations of two consecutive springs. It can be clearly seen that the equations of motion in Eq.~\eqref{MDDimer_EqsMotion_Elong} resemble the displacement equations of motion of the stiffness dimer in Eq.~\eqref{SDDimer_EqsMotion}, and thus again map to the SSH model with onsite potential. Upon substituting plane wave solutions, we will get the deformation dynamical matrix that exhibits chiral symmetry, thus removing the ambiguity of defining the topological phase. Furthermore, for a finite chain, the deformation framework requires an even number of springs and fixed boundaries for chiral symmetry~\cite{allein2023strain}. This translates to a lattice with an odd number of particles with free boundaries in the displacement coordinates. Since the finite lattice preserves the mirror symmetry, the boundary states have the same localization profiles on both ends as in the case of the stiffness dimer.

We have seen that the two canonical models discussed so far use spring connections that store a potential energy of $\frac{1}{2} k (u_1 - u_2)^2$ for any two degrees of freedom, $u_1$ and $u_2$. To generalize such models further with different forms of potential energy and explore hidden symmetries within the deformation framework, we consider a general quadratic potential:
\begin{equation*}
U = \frac{1}{2} k \left(u_1^2 + \chi u_2^2 + 2\psi u_1 u_2 \right),
\end{equation*}
where $k$ is the spring constant, and $\chi$ and $\psi$ are non-dimensional coefficients that depend on the type of coupling. 
To represent a dimer model in deformation coordinates, one requires $\chi = \psi^2$, which implies the following form of the potential: 
$U = \frac{1}{2} k \left(u_1 + \psi u_2 \right)^2 = \frac{1}{2} k e^2 $, where $e$ is the deformation (see SM for the derivation).
However, for dimer models to exhibit chiral symmetry in the deformation coordinates, we also require $\psi = \pm 1$. 
This yields only two possible types of spring potential, $\frac{1}{2} k (u_1 \pm u_2)^2$, that can be used to construct a dimer chain with hidden chiral symmetry in deformation coordinates.

Based on these constraints, we propose a model inspired by the mass dimer but featuring two distinct types of spring connections, as shown in Fig.~\ref{Fig1}C. Although all spring constants are denoted by $k$, the manner in which they store potential energy allows for classification into two types. The first type is a conventional spring connection that stores a potential energy of $\frac{1}{2} k (u_1 - u_2)^2$. In contrast, the second type of spring connection yields a potential energy of $\frac{1}{2} k (u_1 + u_2)^2$. Due to the differing deformations in the two types of spring connections, we term this model the ``deformation mass dimer." 
The equations of motion for the $n$th unit cell can thus be expressed as
\begin{equation}
    \begin{aligned}
        m_1 \Ddot{u}_{1,n} &= -k (u_{1,n} - u_{2,n}) - k(u_{2,n-1} + u_{1,n}) \\
        m_2 \Ddot{u}_{2,n} &= +k (u_{1,n} - u_{2,n}) - k(u_{2,n} + u_{1,n+1}).
    \end{aligned}
    \label{EMDDimer_EqsMotion}
\end{equation}
Upon substituting plane wave solutions, we get the non-dimensional bulk displacement dynamical matrix $\mathcal{D}_u(P,q)$ that has the following Pauli components: $c_0 = 1+1/P$, $c_1 = (\cos{q}-1)/\sqrt{P}$, $c_2 = \sin{q}/\sqrt{P}$, and $c_3 = 1-1/P$, where $P \coloneqq m_2/m_1$.

Similar to the mass dimer, the dynamical matrix lacks chiral symmetry in the displacement framework. However, unlike the mass dimer, the deformation coordinates for the $n$th unit cell of this model are given as
\begin{equation*}
    \begin{aligned}
        e_{1,n} &= u_{1,n} - u_{2,n} \\
        e_{2,n} &= u_{2,n} + u_{1,n+1}
    \end{aligned}
\end{equation*}
and the corresponding equations of motion are
\begin{equation}
    \begin{aligned}
        \frac{1}{k} \Ddot{e}_{1,n} &= -\frac{1}{m_2} (e_{1,n} - e_{2,n}) - \frac{1}{m_1} (e_{2,n-1} + e_{1,n}) \\
        \frac{1}{k} \Ddot{e}_{2,n} &= +\frac{1}{m_2} (e_{1,n} - e_{2,n}) - \frac{1}{m_1} (e_{2,n} + e_{1,n+1}).
    \end{aligned}
    \label{EMDDimer_EqsMotion_Elong}
\end{equation}
Now, upon substituting plane wave solutions, we get the deformation dynamical matrix as
\begin{equation*}
    \mathcal{D}_e(P,q) = \frac{1}{P}
    \begin{bmatrix}
        1+P & -(1-P \mathrm{e}^{-iq}) \\
        -(1-P \mathrm{e}^{iq}) & 1+P
    \end{bmatrix},
\end{equation*}
which exhibits chiral symmetry. Therefore, this deformation mass dimer only reveals the hidden symmetry in deformation coordinates. Interestingly, the eigenvalue problem can be mapped to a 1D SSH chain with an onsite potential and alternating positive and negative hopping~\cite{asboth2016short}. Consequently, the dispersion curve has a band gap at $q=0$ instead of $q=1$  as in the canonical models, revealing non-trivial topology for $P>1$.

Furthermore, we observe that a finite lattice with \textit{odd} particles under free boundary conditions exhibits boundary states on both ends for $P>1$. In contrast to previous canonical models, this finite lattice lacks mirror symmetry due to the presence of two different types of spring connections, resulting in distinct localization profiles of the boundary states on different edges shown in Fig.~\ref{Fig1}C. The hidden mirror symmetry is unveiled in the deformation coordinates discussed here, resulting in localization profiles that closely resemble those of the 1D SSH model with both positive and negative hopping (see Section 1.5.2 in~\cite{asboth2016short}). 
%

\textbf{Mixed spinner lattice}
\begin{figure*}[!]
    \centering
    \includegraphics[width=1.0\textwidth]{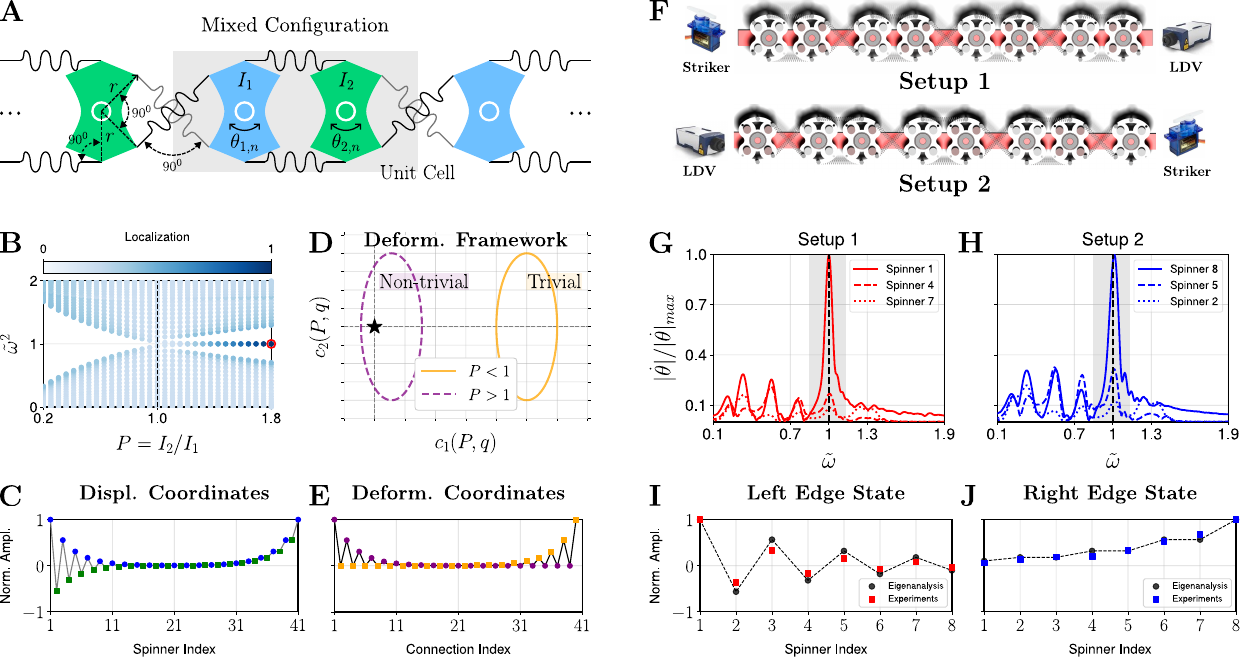}
    \caption{\textbf{Mixed spinner lattice} --
    \textbf{(A)} Spinner chain with parallel and cross-connections. The highlighted unit cell comprises spinners with distinct moments of inertia and pre-stretched springs of stiffness $k/2$. 
    \textbf{(B)} Finite spectrum of $41$-spinner chain with free boundaries as a function of $P$. Localized states emerge inside the bandgap for $P>1$. Here, ${\Tilde {\omega} \equiv \omega/\omega_0}$ is the angular frequency normalized with respect to the mid-gap angular frequency $\omega_0 = \sqrt{1+1/P}$. 
    \textbf{(C)} Boundary state at ${P=1.8}$ show distinct localization profiles at opposite edges.
    \textbf{(D)} Winding number evaluated in the deformation framework. The parametric loop excludes and includes the origin in the trivial and non-trivial case, respectively. 
    \textbf{(E)} The boundary state of a chiral symmetric system at ${P=1.8}$ is revealed in deformation coordinates.
    \textbf{(F)} Experimental setup 1 and setup 2 of spinner chains for probing the left and right boundary states, respectively.
    \textbf{(G, H)} Normalized Fourier spectra of the measured velocities at different spinners along the chain. Spinner counting starts from the left to the right. The shaded region indicates the bandgap. 
    \textbf{(I, J)}  Measured boundary state profiles for ${\Tilde{\omega} = 1}$ compared against eigenanalysis.}
    \label{Fig2}
\end{figure*}

To realize the deformation mass dimer, we construct a spinner lattice as illustrated in Fig.~\ref{Fig2}A. The lattice consists of spinners with two distinct rotational inertias ($I_1$ and $I_2$), arranged alternately along the chain. The spinners can be interconnected using various types of spring connections, such as parallel, cross, and colinear (see details in SM). However, to realize the deformation mass dimer, we utilize a mixed configuration featuring alternating parallel and cross-connections. Colinear connection, previously used to model magnetic spinners~\cite{qian2018topology}, cannot be employed here since its potential energy can not be represented in the following form: $\frac{1}{2} k (u_1 \pm u_2)^2$ as required for the deformation formulation. Nevertheless, they can be used to extend the 1D mixed chain into a 2D configuration, as discussed in later sections.

Furthermore, in our mixed spinner 1D chain, the cross-connections form a 90-degree angle with each other and with the position vector of the connecting points. This arrangement ensures the required form of the potential and maintains the same initial length of all springs (with a spring constant of $k/2$) for experimental convenience (see details in SM). Therefore, the equations of motion governing the angular rotations $\theta$ of the two spinners within the $n$th unit cell are expressed as
\begin{equation}
    \begin{aligned}
        I_1 \Ddot{\theta}_{1,n} &= - kr^2(\theta_{1,n} - \theta_{2,n}) - kr^2(\theta_{2,n-1} + \theta_{1,n}) \\
        I_2 \Ddot{\theta}_{2,n} &= + kr^2(\theta_{1,n} - \theta_{2,n}) - kr^2 (\theta_{2,n} + \theta_{1,n+1}),
    \end{aligned}
    \label{RotEqsMotion}
\end{equation}
where $r$ is the radial distance of the connection points, $\theta_{1,n}$ and $\theta_{2,n}$ represent the angular rotation of the first and second spinners within the $n$th unit cell, respectively, and the overdot denotes the time derivative. Thus, the mixed spinner system mimics the dynamics of the deformation mass dimer in Eq.~\eqref{EMDDimer_EqsMotion}. 

We seek plane wave solutions of the form ${[\theta_{1,n}, \theta_{2,n}] = [\theta_1(q), \theta_2(q)] \mathrm{e}^{i(q n - \Omega t)}}$, where \( q \in [-\pi,\pi] \) and $\Omega$ denote the non-dimensional wavenumber and angular frequency, respectively. This leads to the eigenvalue problem: $\mathcal{D}_u(P,q) \boldsymbol{\theta}(q) = \omega^2 \boldsymbol{\theta}(q)$, where $\boldsymbol{\theta}(q) \coloneq [\theta_1(q)/\sqrt{I_1}, \theta_2(q)/\sqrt{I_2}]$, $\omega \coloneq \Omega/\sqrt{kr^2/I_1}$, $P \coloneq I_2/I_1$, and the displacement dynamical matrix is given by
\begin{equation*}
    \mathcal{D}_u(P,q) \coloneq
    \begin{bmatrix}
        2 & -\frac{1}{\sqrt{P}}(1 - \mathrm{e}^{-i q}) \\
        -\frac{1}{\sqrt{P}}(1 - \mathrm{e}^{i q}) & 2/P
    \end{bmatrix}.
\end{equation*}
This is the bulk matrix for the deformation mass dimer discussed in the context of Fig.~\ref{Fig1}C. The matrix lacks chiral symmetry.

In a finite lattice with an odd number of spinners (fractional unit cells) and free boundary conditions, localized states emerge at the mid-gap squared frequency for \( P>1 \), as depicted in Fig.~\ref{Fig2}B (see the 1D localization parameter in \textit{Materials and Methods}). Here, \( P>1 \) signifies that the system terminates with lower-inertia spinners on both ends. In Fig.~\ref{Fig2}C, we present the boundary state profiles for \( P=1.8 \). Though some aspects of this scenario resemble a known topological transition with the emergence of topological boundary states, the necessary symmetry enabling topological characterization and bulk-boundary correspondence remains unclear.
 
To reveal the hidden topology, we consider the dynamics of our system in deformation coordinates defined as:
$$
    \begin{aligned}
        e_{1,n} &= \theta_{1,n} - \theta_{2,n} \\
        e_{2,n} &= \theta_{2,n} + \theta_{1,n+1}.
    \end{aligned}
$$
Upon rearranging Eq.~\eqref{RotEqsMotion} and seeking plane-wave solutions ${[e_{1,n}, e_{2,n}] = [e_1, e_2] \mathrm{e}^{i(q n - \Omega t)}}$
we obtain the eigenvalue problem: ${\mathcal{D}_e(P,q) \boldsymbol{e}(q) = \omega^2 \boldsymbol{e}(q)}$, 
where ${\boldsymbol{e}(q)  \coloneq [e_1(q), e_2(q)]}$ and the deformation dynamical matrix is defined as 
$$
    \mathcal{D}_e(P,q)  \coloneq \frac{1}{P}
    \begin{bmatrix}
        1+P & -(1 - P \mathrm{e}^{-i q}) \\ 
        -(1 - P \mathrm{e}^{i q}) & 1+P
    \end{bmatrix}.
    \label{Struct_Ds}
$$
When expressed in Pauli representation, ${\mathcal{D}_e(P,q) = d_0 \mathcal{I} + d_1 \sigma_1 + d_2 \sigma_2}$, with ${d_0 = 1+1/P}$, ${d_1 = \cos{q} - 1/P }$, and ${d_2 = \sin{q}}$. This implies the presence of chiral symmetry. The winding number ($\nu_e$) reveals the topological phase distinction for ${P<1}$ ($\nu_e = 0$) and ${P>1}$ ($\nu_e = 1$) denoting trivial and non-trivial phases, respectively, as shown in Fig.~\ref{Fig2}D.

Therefore, the system exhibits bulk-boundary correspondence in the deformation coordinates. In Fig.~\ref{Fig2}E, we depict the boundary state profiles of Fig.~\ref{Fig2}C in the transformed coordinates. Remarkably, the chiral nature of the boundary states is revealed in that alternating particles are at rest, as in the canonical stiffness dimer, but with alternating positive and negative couplings. Moreover, the mirror symmetry of the lattice is revealed, leading to identical localization profiles for both ends. 

Next, we experimentally validate the presence of such topological states. We choose an even number of spinners for the experiments to detect a boundary state on one end at a time for a short chain. Therefore, we construct a spinner chain comprising eight spinners and conduct two sets of experiments, as shown in Fig.~\ref{Fig2}F, to detect topological boundary states on the left and right ends separately. See \textit{Materials and Methods} for more details on the fabrication and data acquisition.

In Figs.~\ref{Fig2}G-H, we present the Fourier spectrum of the measured velocities of different spinners. We observe high amplitude peaks inside the bandgap for spinner 1 and spinner 8, indicating the presence of boundary states. In Figs.~\ref{Fig2}I-J, we plot the spatial profiles corresponding to the high amplitude peaks. We find that both left and right boundary states match well with eigenanalysis. The slight mismatch could be attributed to several effects, such as spring misalignment, friction, and nonlinearity. Therefore, we have demonstrated boundary states in a spinner chain with free boundaries with hidden topology in deformation coordinates, and such boundary states have different profiles at different edges.
\begin{figure*}[!]
    \centering
    \includegraphics[width=1.0\textwidth]{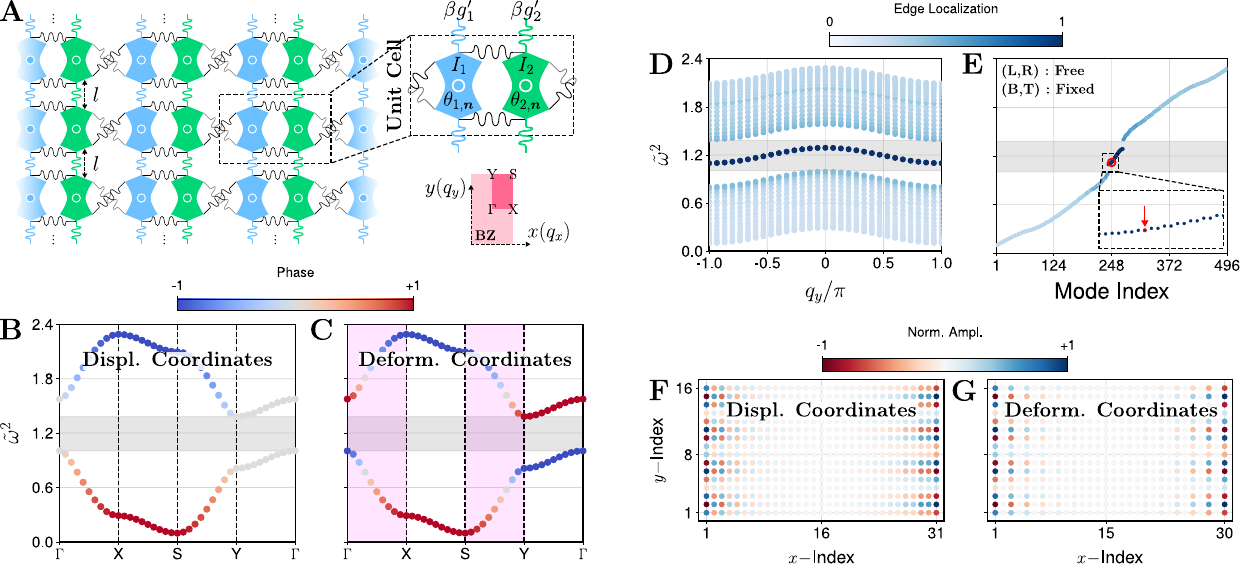}
    \caption{\textbf{2D mixed spinner lattice} --
    \textbf{(A)} 2D lattice with identical 1D chains coupled along the $y$-axis with spacing $l$. The inset shows the unit cell with effective stiffness coefficients and inertia.
    Dispersion diagram with a complete bandgap (grey-shade) obtained from the displacement \textbf{(B)} and the deformation \textbf{(C)} dynamical matrix at $(P,\beta)=(1.8,0.075)$. The color bar denotes the relative phase between the eigenvector components.
    The spectrum of \textbf{(D)} a ribbon finite in the $x$-direction and \textbf{(E)} a finite 2D lattice with free left-right boundaries. For the latter, the bottom-top boundaries are fixed, and the inset shows a snippet of the distribution of edge states in the bandgap. The color bar denotes the edge localization, indicating the presence of edge states inside the band gap.
    \textbf{(F, G)} An edge state in the finite 2D lattice in displacement and deformation coordinates.}
    \label{Fig3}
\end{figure*}

\textbf{2D extension}
%
%

We now extend our spinner lattice to a 2D configuration. This extension is non-trivial because it (1) allows for the realization of propagating waves with profiles that differ at the two opposite finite edges and (2) enables a further generalization of the deformation framework beyond 1D lattices. We construct a 2D lattice composed of spinners by stacking 1D chains ($x$-axis) separated by a distance $l$ in the other extended direction ($y$-axis) as shown by an example in Fig.~\ref{Fig3}A. Spinners with rotational inertia $I_1$ and $I_2$ are coupled to their nearest neighboring chains with effective spring constants $\beta g_1'$ and $\beta g_2'$, respectively. Here, $\beta$ is a scale factor that indicates the strength of inter-chain coupling. The generalized equations of motion governing the angular rotations $\theta$ of the two spinners within the $\boldsymbol{n}$th unit cell (${\boldsymbol{n} = n_x \hat{\boldsymbol{e}}_x + n_y \hat{\boldsymbol{e}}_y}$) are given as
\begin{equation}
    \begin{aligned}
        &I_1 \Ddot{\theta}_{1,\boldsymbol{n}} = -kr^2\left[\theta_{2(n_x-1,n_y)} + 2\theta_{1(n_x,n_y)} - \theta_{2(n_x,n_y)}\right] \\
        &- \hat{k}_1 r^2 \left[\psi \theta_{1(n_x,n_y-1)} + (1+\chi) \theta_{1(n_x,n_y)} + \psi \theta_{1(n_x,n_y+1)}\right] \\[5pt]
        &I_2 \Ddot{\theta}_{2,\boldsymbol{n}} = -kr^2 \left[-\theta_{1(n_x,n_y)} + 2\theta_{2(n_x,n_y)} + \theta_{1(n_x+1,n_y)}\right] \\
        &- \hat{k}_2 r^2 \left[\psi \theta_{2(n_x,n_y-1)} + (1+\chi) \theta_{2(n_x,n_y)} + \psi \theta_{2(n_x,n_y+1)}\right],
    \end{aligned}
    \label{2DRotEqsMotion_General}
\end{equation}
where $\hat{k}_{1/2} = \beta g_{1/2}'$ and the non-dimensional coefficients, $\psi$ and $\chi$, depend upon the type of spring connection along $y$-axis. For example, we can use a parallel or cross-spring connection as employed in the 1D configuration, where the coefficients $(\chi, \psi)$ are $(1, -1)$ and $(1, 1)$, respectively. Alternatively, we can use the colinear connection with the coefficients $\left[\chi=1,\psi=r/(r+l)\right]$ along the $y$-axis (as shown in Fig.~\ref{Fig3}A), which could not be used for the 1D setting earlier.

In Eq.~\eqref{2DRotEqsMotion_General}, we can once again define $P \coloneq I_2/I_1$ and also $P' \coloneq g'_2/g'_1$. We then seek the plane wave solutions of the form ${[\theta_{1, \boldsymbol{n}},\theta_{2, \boldsymbol{n}}] = [\theta_1(\boldsymbol{q}), \theta_2(\boldsymbol{q})]\mathrm{e}^{i[\boldsymbol{q} \cdot \boldsymbol{n} - \Omega t]}}$, where ${\boldsymbol{q} = q_x \hat{\boldsymbol{e}}_x + q_y \hat{\boldsymbol{e}}_y}$. We thus obtain the eigenvalue problem: ${\mathcal{D}_u(P,\beta,\boldsymbol{q}) \boldsymbol{\theta}(\boldsymbol{q}) = \omega^2 \boldsymbol{\theta}(\boldsymbol{q})}$, where ${ \boldsymbol{\theta}(\boldsymbol{q}) \coloneq [\theta_1(\boldsymbol{q}) /\sqrt{I_1},\theta_2(\boldsymbol{q})/\sqrt{I_2}]}$, and the displacement dynamical matrix is given by
\begin{equation*}
    \mathcal{D}_u(P,\beta,\boldsymbol{q}) \coloneq \mathcal{D}_u(P,q_x) + \beta f(q_y)
    \begin{bmatrix}
        1 & 0  \\
        0 & P'/P
    \end{bmatrix}
    \label{2DStruct_Du_General}
\end{equation*}
with $f(q_y) = \left(g'_1/k\right) \left[(1+\chi)+2\psi\cos(q_y) \right]$. We observe that using different spring connections along the $y$-direction yields different $f(q_y)$; however, they do not alter the essential physics discussed hereafter.
From hereon, we employ the colinear connection for its experimental convenience and to demonstrate the versatility of our design (the cases with parallel and cross-connections along the $y$-axis have been discussed in SM for completeness). Without any loss of generality, we assume $g'_1 = k \left(1+l/r\right)$ and $l = r$ for simplicity. This then yields $f(q_y) = 2+4\cos^2(q_y/2)$. Once again, the dynamical matrix $\mathcal{D}_u(P,\beta,\boldsymbol{q})$ lacks chiral and inversion symmetries, making topological characterization unclear.

Interestingly, the 2D displacement dynamical matrix with the specific fine-tuning
\begin{equation*}
    P' = P
\end{equation*}
can be interpreted as a 1D displacement dynamical matrix with a $q_y$-dependent spectral shift, i.e., ${\mathcal{D}_u(P,\beta,\boldsymbol{q}) = \mathcal{D}_u(P,q_x) + \beta f(q_y)\mathcal{I}}$. This 2D extension is akin to the tight-binding Hamiltonian of the dimerized square lattice discussed in earlier works~\cite{hughes2011inversion, chen2018two}. However, these works involved inversion symmetric systems, which our system lacks in displacement coordinates.

Now, to represent the displacement dynamical matrix in deformation coordinates and to reveal the hidden symmetry, we employ the following deformation coordinates as per the horizontal 1D chains: ${ e_{1,\boldsymbol{n}} = \theta_{1(n_x,n_y)} - \theta_{2(n_x,n_y)} }$ and ${e_{2,\boldsymbol{n}} = \theta_{1(n_x,n_y)} + \theta_{2(n_x+1,n_y)}}$ and substitute the plane wave solution. This results in the eigenvalue equation ${\mathcal{D}_e(P,\beta,\boldsymbol{q}) \boldsymbol{e} (\boldsymbol{q}) = \omega^2 \boldsymbol{e}(\boldsymbol{q})}$, in which the deformation dynamical matrix can be recast as $$\mathcal{D}_e(P,\beta,\boldsymbol{q}) = \mathcal{D}_e(P,q_x) + \beta f(q_y) \mathcal{I}.$$ This implies that the 2D deformation dynamical matrix is a 1D deformation dynamical matrix with a $q_y$-dependent spectral shift. Moreover, it follows the chiral-like symmetry
\begin{equation*}
     \sigma_3 \mathcal{D}_e(P,\beta,\boldsymbol{q}) \sigma_3 + \mathcal{D}_e(P,\beta,\boldsymbol{q})
     = 2 \left[1 + \frac{1}{P} + \beta f(q_y) \right] \mathcal{I}
\end{equation*}
along the $x$-axis where the right-hand side in the equation is not a constant but $q_y$-dependent. 

We can now characterize the topology using the vectored Zak phase~\cite{obana2019topological}
\begin{equation*}
    \boldsymbol{Z}_e \coloneq \frac{1}{i} \sum_{j=x,y} \int_{-\pi}^{+\pi} \bra{e(\boldsymbol{q})} \frac{\partial}{\partial q_j} \ket{e(\boldsymbol{q})} dq_j \hat{\boldsymbol{\mathrm{e}}}_j,
\end{equation*}
which deduces to 
\begin{equation*}
    \boldsymbol{Z}_e = (Z_e^x,Z_e^y) =
    \begin{cases}
       (0,0),~P<1 \\
       (\pi,0),~P>1
    \end{cases}
\end{equation*}
demonstrating the onset of non-trivial topological configuration along the $x$-axis for ${P>1}$. Importantly, this is irrespective of $\beta$ as long as the condition $P' = P$ is enforced on the spring constants. 
The quantized values of $Z_e^x$ stem from the fact that the deformation dynamical matrix exhibits an inversion symmetry along the $x$-axis as follows
\begin{equation*}
    \sigma_1 \mathcal{D}_e \left(P,\beta,\boldsymbol{q}=(q_x,q_y)\right) \sigma_1 = \mathcal{D}_e \left(P,\beta,\boldsymbol{q}=(-q_x,q_y)\right).
    \label{InvSymm2D}
\end{equation*}
We plot the dispersion curve from the displacement and the deformation dynamical matrix in Figs.~\ref{Fig3}B-C for the non-trivial case $(P,\beta)=(1.8,0.075)$  First, we observe a complete band gap, which exists as long as $\beta < \abs{1 - 1/P}/2$.  In the $\Gamma X$ direction, we observe that the two dispersion curves are symmetric around a mid-frequency.  By also plotting the phases of eigenvectors, we observe the phase reversal of a branch from $\Gamma$ to $X$ in deformation coordinates, indicating non-trivial topology. $\Gamma$ to $Y$ direction in deformation coordinates, however, shows no such phase reversal, therefore complying with trivial topology along the $y$-axis. 

In Fig.~\ref{Fig3}D, we plot the spectrum of a ribbon consisting of 41 spinners, with free boundaries along the $x$-axis and Floquet boundary conditions along the $y$-axis (see the 2D localization parameter in \textit{Materials and Methods}). We observe localized states on the left and right boundaries of the ribbon in the band gap for all values of $q_y \in [-\pi, \pi]$. We then consider a finite structure with 31 and 16 spinners along the $x$ and $y$-axes, respectively. In Fig.~\ref{Fig3}E, we plot the spectrum of the finite 2D system with free (fixed) left-right (top-bottom) boundaries. We observe multiple edge states in the band gap. These edge states exhibit different localization profiles at the left and right edges, as shown in Fig.~\ref{Fig3}F, which is a hallmark of our mixed spinner lattice. Upon transformation to deformation coordinates, the chiral-like symmetry is revealed, as shown in Fig.~\ref{Fig3}G, where every alternate particle is at rest, similar to the 1D chain in the deformation framework. The lattice also reveals the mirror symmetry along the $x$-axis in the deformation framework, similar to the 1D chain.

The nature of eigenvectors for a finite system can be understood by looking at the finite dynamical matrix. For a lattice with $N$ and $M$ spinners along the $x$ and $y$-axes, respectively, the displacement dynamical matrix of size $NM \times NM$ is expressed as follows:
\begin{equation*}
    D_u^{\text{finite}} = D_u^x \oplus \beta D^y,
\end{equation*}
where $\oplus$ denotes Kronecker sum, $D_u^x \in \mathbb{R}^{N \times N}$ is the finite 1D displacement dynamical matrix of the mixed configuration (along the $x$-axis) and $D^y \in \mathbb{R}^{M \times M}$ is the finite coupling matrix characterizing the coupling between stacked 1D chains along the $y$-axis. This separability is achieved by enforcing $P'=P$ (see SM for the detailed derivation). Therefore, the displacement eigenvectors of the finite system can be expressed as
\begin{equation}
    \boldsymbol{u}^{\text{2D}}_{i,j} = \boldsymbol{p}_i \otimes \boldsymbol{u}_j,
    \label{Finite2DEigVec}
\end{equation}
for the corresponding eigenfrequency $\omega_{i,j}^2 = \beta \mu^2_i + \omega_{u(j)}^2$, where we have the following 1D eigenvalue equations: $D^y \boldsymbol{p}_i = \mu_i^2 \boldsymbol{p}_i$ and $D_u^x \boldsymbol{u}_j = \omega_{u(j)}^2 \boldsymbol{u}_j$ for all $i,j$~\cite{coutant2020robustness,coutant2021topological}. 
Since the 1D chain has two edge states, $\boldsymbol{b}_{\pm}$, in our 2D lattice, we will have $2M$ edge states of the form $\boldsymbol{p}_i \otimes \boldsymbol{b}_{\pm}$ for all $i \in \{1,2,...,M\}$. 

Similarly, in the deformation coordinates, the dynamical matrix of size $(N-1)M \times (N-1)M$ is expressed as follows:
\begin{equation*}
    D_e^{\text{finite}} = D_e^x \oplus \beta D^y,
\end{equation*}
where $D_e^x \in \mathbb{R}^{(N-1) \times (N-1)}$ is the finite 1D deformation dynamical matrix of the mixed configuration (see SM for details). The deformation eigenvectors of the finite system are expressed as
\begin{equation}
    \boldsymbol{e}^{\text{2D}}_{i,j} = \boldsymbol{p}_i \otimes \boldsymbol{e}_j,
    \label{Finite2DEigVecStr}
\end{equation}
for the corresponding eigenfrequency $\omega_{i,j}^2 = \beta \mu^2_i + \omega_{e(j)}^2$, where we have the following 1D eigenvalue equation, $D_e^x \boldsymbol{e}_j = \omega_{e(j)}^2 \boldsymbol{e}_j$ for all $j$.
Thus, $D_u^{\text{finite}}$ and $D_e^{\text{finite}}$ form a quasi-isospectral pair, with $(N-1)M$ out of $NM$ eigenvalues of $D_u^{\text{finite}}$ mapping to the eigenspectrum of $D_e^{\text{finite}}$.
The remaining $M$ eigenvalues of $D_u^{\text{finite}}$ are related to the zero modes of the 1D free-free chain. In 2D, these are spectrally shifted, such that $\omega_{i}^2 = \beta \mu^2_i $, and are not part of the deformation dynamical matrix.  
Moreover, the eigenvectors for the edge states in the 2D lattice are easily related to those in the 1D lattice. Therefore, the chiral-like nature of edge states on both edges of the 2D lattice is naturally revealed in deformation coordinates, as shown in Fig.~\ref{Fig3}G.

\textbf{Robustness to disorder}
\begin{figure*}[!]
    \centering
    \includegraphics[width=1.0\textwidth]{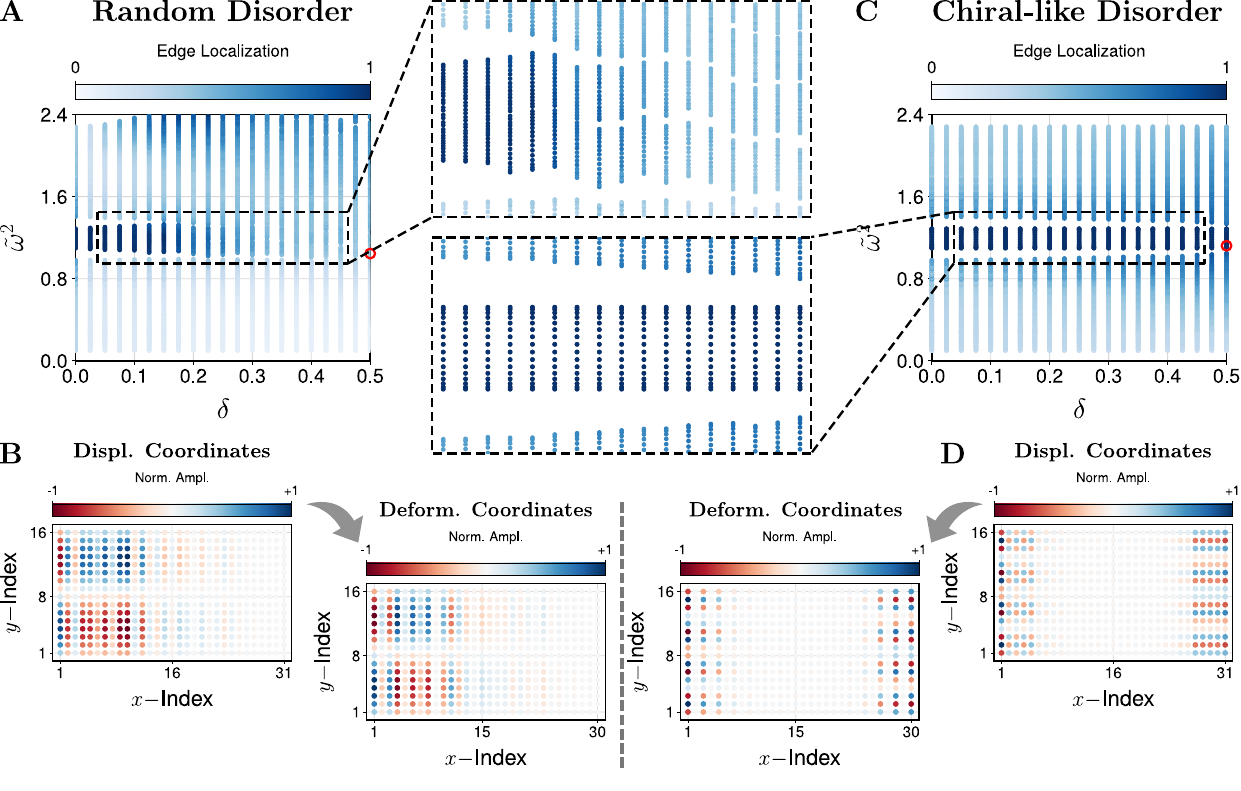}
    \caption{\textbf{Disorder analysis} --
    Variation of the averaged (over 50 realizations) eigenspectrum of the lattice of size $N \times M = 31 \times 16$ with free left-right and fixed top-bottom boundaries with respect to \textbf{(A)} random and \textbf{(C)} chiral-like disorders (symmetry-preserving). $\delta$ denotes the corresponding disorder strength. The zoomed plots show that the edge states are unstable and stable against increasing disorder strength. 
    An edge state sample in displacement and deformation coordinates for \textbf{(B)} random and \textbf{(D)} chiral-like disorders at $\delta=0.5$  The latter case demonstrates the unaffected localization and chiral profile of edge states in displacement and deformation coordinates, respectively.}
    \label{Fig4}
\end{figure*}

The underlying symmetry in the deformation coordinates also helps to discover disordered lattice configurations that lead to the topological edge states that are robust.
We define the 2D chiral-like operator as $\Pi^{\text{2D}} \coloneq \mathcal{I}_M \otimes \Pi^x$, where ${\Pi^x \in \mathbb{R}^{(N-1) \times (N-1)} = diag(+1,-1,+1,...)}$ is the finite 1D chiral operator. $\Pi^{\text{2D}}$ anti-commutes with the global deformation dynamical matrix as follows
\begin{equation*}
    \Pi^{\text{2D}} D_e^{\text{finite}} \Pi^{\text{2D}} + D_e^{\text{finite}} = 2(1+1/P)\mathcal{I}_N \oplus \beta D^y.
\end{equation*}
The $\Pi^x$ term in $\Pi^{\text{2D}}$ acts only on the $D_e^x$ term in $D_e^{\text{finite}}$ and thus protects all edge states in the system because $D_e^x$ exhibits chiral symmetry~\cite{shi2021disorder}.

We perform a generalized study to determine the spectrum by introducing disorder in the system along the $x$-axis, i.e., quantities remain uniform along the $y$-axis. First, we take $N$ spinners with disordered moments of inertia such that
\begin{equation*}
    \tilde{I}_p = I_p^0 (1 + \delta \lambda_p) ~\forall~ p \in \{1,2,...,N\},
\end{equation*}
where ${I_{\text{odd}/\text{even}}^0 = I_{1/2}}$, $\lambda_p$ takes random values in $[-1,1]$, and $\delta$ indicates the strength of the disorder. Then, we consider interconnecting springs in mixed configuration with spring constants given as
\begin{equation*}
    \tilde{k}_q = k (1 + \delta \eta_q) ~\forall~ q \in \{1,2,...,N-1\},
\end{equation*}
where $\eta_q \in [-1,1]$. Second, we stack identical copies of the disordered 1D chain along the $y$-axis coupled by pre-stretched springs with disordered effective spring constants given as $\beta \tilde{g}'_r = \beta g_r^0 (1 + \delta \phi_r) ~\forall~r \in \{1,2,...,N\}$, where $g_{\text{odd}/\text{even}}^0 = g'_{1/2}$ and $\phi_r \in [-1,1]$.

Let $\Tilde {D}_e^{\text{finite}}$ denote the disordered global deformation dynamical matrix. To satisfy the chiral-like symmetry, i.e., $\Pi^{\text{2D}} \tilde{D}_e^{\text{finite}} \Pi^{\text{2D}} + \tilde{D}_e^{\text{finite}} = 2(1+1/P)\mathcal{I}_N \oplus D^y$, the first condition stems form the separability aspect, i.e., $\Tilde {D}_e^{\text{finite}}$ must be of the form
\begin{equation*}
    \tilde{D}_e^{\text{finite}} = \tilde{D}_e^x \oplus \beta D^y,
\end{equation*}
where $\tilde{D}_e^x \in \mathbb{R}^{(N-1) \times (N-1)}$ is the disordered finite 1D deformation dynamical matrix. This imposes the first constraint
\begin{equation*}
    \tilde{g}_i/\tilde{I}_i = g_i^0/I_i^0 \implies \phi_i = \lambda_i
\end{equation*}
for all $i \in \{1,2,3,...N\}.$

The second constraint stems from the action of $\Pi^x$ in the chiral operator on $\tilde{D}_e^x$ in $\tilde{D}_e^{\text{finite}}$ as follows:
${\Pi^x \tilde{D}_e^x \Pi^x + \tilde{D}_e^x = 2(1+1/P)\mathcal{I}_N }$. This
imposes $\tilde{k}_i (1/\tilde{I}_i + 1/\tilde{I}_{i+1}) = k (1/I_i^0 + 1/I_{i+1}^0)$, and therefore, $\eta_i$ can be determined as 
\begin{equation*}
    \eta_i (\lambda_i, \lambda_{i+1}) = \frac{\lambda_i h_i + \lambda_{i+1} \left( h_{i+1}/P \right)}{h_i + \left( h_{i+1}/P \right)}
\end{equation*}
for all $i \in \{1,2,3,...N-1\}$, where $h_x = 1/\left(1+\delta \lambda_x \right)$. Hence, we have derived the conditions necessary to preserve the chiral-like symmetry in a disordered spinner system.

Figures~\ref{Fig4}A and \ref{Fig4}C depict the variation of the eigenspectrum of the disordered finite system of size $N \times M = 31 \times 16$ with respect to random (random values of $\lambda_p$, $\eta_q$, and $\phi_r$ for all $p,q$ and $r$) and chiral-like symmetry preserving disorders.  The random disorders destabilize the edge states, with their localization properties being strongly affected by the strength of the disorder, as seen in Fig.~\ref{Fig4}B for $\delta=0.5$.  On the other hand, disorders respecting the underlying chiral-like symmetry do not affect the edge states, and they remain pinned to their frequencies, exhibiting extreme robustness.  An edge state sample in Fig.~\ref{Fig4}D for $\delta=0.5$ confirms that the localization profiles are intact, and the deformation coordinates reveal the hidden chiral profile like in 1D chains. 

\textbf{Conclusions and outlook}

In summary, we have introduced a prototypical spring-mass model incorporating distinct spring connectivities and demonstrated both theoretically and experimentally the presence of topological edge states with different profiles at opposite ends. The topological nature of the lattice is uncovered through deformation coordinates, which simultaneously reveal the underlying chiral and mirror symmetries. We further extend this model to a two-dimensional setting, where precise fine-tuning is necessary to apply the deformation framework and highlight the topologically robust nature of the waves propagating along opposite lattice edges.

Looking ahead, the design space provided by our spinner-based models offers a valuable framework for exploring various tight-binding 2D models. Additionally, this approach opens opportunities for investigating novel classes of higher-order topological lattices~\cite{serra2018observation}, with diverse boundary conditions and localization profiles based on hidden symmetries.


%
\textcolor{red}{\textbf{METHODS}}

\textbf{Setup fabrication}

The spinners, along with the bearings ($608$RS), are mounted on a 3D-printed ABS base. Each spinner is individually 3D-printed using tough PLA (mass density ${\rho = 1.20}$ g/cm$^3$). Four slots are incorporated into each spinner for metal inserts.  Aluminum (density $\rho = 2.70$ g/cm$^3$) and copper (density $\rho = 8.94$ g/cm$^3$) cylindrical stubs of the following dimensions: diameter $\phi = 1$ cm and height $h = 2$ cm are alternately utilized in the spinners to impart different moments of inertia.  Additional holes are provided in the spinners for M$2$ screws for attaching tension springs.  The spinners are coupled to their nearest neighbors by alternating parallel and cross-connections.  The positioning of the spinners and spring connections is meticulously arranged to maintain ($1$) uniform pre-stretch in the springs throughout the setup and ($2$) the desired angular orientation of the cross-connections.

\textbf{Setup parameters}

In order to normalize frequencies obtained from the finite chains, the natural frequency of angular oscillations of a single Aluminum spinner ($f_{\text{Al}} = \sqrt{kr^2/I_1}/2\pi$) is determined through small impact tests. The averaged Fourier spectrum of the time series data obtained from the impacts revealed $f_{\text{Al}} \approx 12.0675$ Hz. Similarly, the natural frequency of angular oscillations of a single copper spinner ($f_{\text{Cu}} = \sqrt{kr^2/I_2}/2\pi$) is obtained as $f_{\text{Cu}} \approx 9.0766$ Hz.  The ratio $f_{\text{Al}}/f_{\text{Cu}} = \sqrt{I_2/I_1} = \sqrt{P}$ then reveals $P \approx 1.7676$ for both experimental setups. Proper bearing lubrication was ensured through WD$40$, leading to averaged damping ratios $\zeta_{\text{Al}} \approx 0.0173$ and $\zeta_{\text{Cu}} = 0.0164$.

\textbf{Modeshape extraction}

The entire assembly is then subjected to automated impulse excitation through a servo motor (Towerpro SG$90$), and the synchronized time series data of all spinners are recorded through a laser Doppler vibrometer (LDV).  The Fourier spectrum of the time series data for all spinners is obtained.  The frequencies are first normalized by $f_{\text{Al}}$ followed by the mid-gap angular frequency $\omega_0 = \sqrt{1+1/P}$. The edge modes are obtained by probing the magnitude and phase of the Fourier data at $\Tilde{\omega} = 1$ for all spinners. Finally, the edge modes are normalized with respect to the maximum magnitude of angular displacements.

\textbf{Localization parameters}

For finite 1D systems, we utilize relative IPR (Inverse Participation Ratio) defined as follows
\begin{equation*}
    \text{IPR}_i^{rel} = \frac{1}{\text{IPR}_{max}}\sum_j \big| v_{i(j)} \big|^4
\end{equation*}
for an eigenvector $\boldsymbol{v}_i$ where $v_{i(j)}$ denotes its $j$th component. This ensures that the boundary states have a relative IPR of unity.

For finite 2D systems, the displacement eigenvectors are obtained by the tensor product of 1D eigenvectors, as shown in Eq.~\eqref{Finite2DEigVec}, the eigenvectors of the dynamical matrix $D^y$ only encode the phase information along the $y$-axis. Therefore, the angular displacements for a single row along the $x$-axis are enough to investigate localization along the left and right edges. Assuming a lattice with $N$ and $M$ spinners along the $x$ and $y$-axes, we consider the truncated displacement eigenvector $\boldsymbol{u}_{i,j}^{\text{2D}}\left(1,N\right)$ obtained by isolating the first $N$ components of the displacement eigenvector $\boldsymbol{u}_{i,j}^{\text{2D}}$ and calculate the relative IPR as follows:
\begin{equation*}
    \text{IPR}_{i,j}^{rel} = \frac{1}{\text{IPR}_{max}}\sum_p \big| u_{i,j(p)}^{\text{2D}}\left(1,N\right) \big|^4
\end{equation*}
where $u_{i,j(p)}^{\text{2D}}\left(1,N\right)$ denotes the $p$th component of the truncated displacement eigenvector. Here, too, the edge states have a relative IPR of unity.

\textbf{Acknowledgments} -- This work is supported by the Science and Engineering Research Board (SERB), India, through the Start-up Research Grant SRG/2022/00166. The authors would like to thank Prof. Fotios Diakonos for his insightful comments and valuable discussions.

\textbf{Author contributions} -- U.V. and R.C. conceived the project. U.V., under the supervision of G.T. and R.C., conceptualized the spinner-based models and developed the theoretical framework. M.I. and U.V. fabricated the setup and conducted the experiments. U.V. analyzed the experimental data. U.V., G.T., and R.C. contributed to the manuscript writing. R.C. supervised the overall project.

\bibliography{myBibliography}
\end{document}